

\input phyzzx
\REF\frenkel{
I.B.Frenkel and N.Yu.Reshetikhin \journal Comm.Math.Phys.&146(92)1.
}
\REF\shiraishi{
J.Shiraishi \journal Phys.Lett.&A171(92)243;\hfill\break
A.Kato, Y.Quano and J.Shiraishi, Preprint UT-618, Univ. Tokyo,1992.
}
\REF\matsuo{
A.Matsuo, Free Field Realization of Quantum Affine Algebra
$U_q(\widehat{sl_2})$, to appear in Phys.Lett.B;
A $q-$deformation of Wakimoto modules, Primary Fields and
Screening Operators, to appear in Comm.Math.Phys.
}
\REF\kimura{
A.Abada, A.H.Bougourzi and M.A.El.Gradechi,
Deformation of the Wakimoto Constraction, Preprint, 1992;\hfill\break
K.Kimura, On Free Boson Representation of the Quantum Affine Algebra
$U_q(\widehat{sl_2})$,
Preprint RIMS-910, Kyoto Univ., 1992.
}
\REF\miwa{
B.Davies, O.Foda, M.Jimbo, T.Miwa and A.Nakayashiki \journal Comm.Math.Phys.
&151\break(93)89;\hfill\break
M.Idzumi, K.Iohara, M.Jimbo, T.Miwa, T.Nakashima and T.Tokihiro
\journal Int.J.Mod.Phys.\break&A8(93)1479;\hfill\break
M.Jimbo, T.Miwa and Y.Ohta \journal Int.J.Mod.Phys.&A8(92)1457;\hfill\break
M.Jimbo, T.Miwa and A.Nakayashiki, Preprint RIMS-904, Kyoto.Univ.,
1992.
}
\REF\bafe{
D.Bernard and G.Felder \journal Comm.Math.Phys.&127(90)145.
}
\REF\frau{
M.Frau, A.Lerda, J.G.McCarthy, J.Sidenius and S.Sciuto\journal
Phys.Lett.&B245(90)453
}
\REF\konno{
H.Konno \journal Phys.Rev.&D45(92)4555.
}
\REF\drinfeld{
V.G.Drinfeld \journal Sov.Math.Dokl.&36(88)212.
}
\REF\fms{
D.Friedan, E.Martinec and S.Shenker \journal Nucl.Phys.&B271(86)93.
}
\REF\pict{
P.Bowknegt, A.Coresole, J.G.McCarthy and P.van Neuwenhuizen \journal
Phys.Rev.&D39\break
(89)2971;\hfill\break
 D.Z.Freedman and K.Pilch \journal Int.J.Mod.Phys.&A4(89)5553;\hfill\break
U.Carow-Watamura, Z.F.Ezawa, K.Harada, A.Tezuka and S.Watamura
\journal Phys.Lett.\hfill\break&B227(89)73; ibid.\journal &B232(89)186;
}
\REF\dotsenko{
Vl.S.Dotsenko and V.A.Fateev \journal Nucl.Phys.&B251(85)691.
}
\REF\flm{
I.Frenkel, J.Lepowsky and A.Meurman, Vertex Operator Algebras
and the Monster, (Academic Press, Boston, 1988).}
\REF\askey{
R.Askey \journal SIAM J.Math.Anal.&11(80)938;\hfill\break
K.Kadell \journal ibid.&19(88)969;\hfill\break
K.Aomoto, 2 Conjectural Formulae For Symmetric A-Type Jackson Integrals,
Preprint Nagoya.Univ., 1993.
}
\REF\malikov{
F.Malikov \journal Int.J.Mod.Phys.&A7(92)623.
}
\REF\fefr{
B.L.Feigin and E.V.Frenkel \journal Lett.Math.Phys.&19(90)307.
}
\REF\mum{
D.Mumford, Tata Lectures on Theta I (Birkh\"{a}ser, Boston, 1983).
}
\REF\verlinde{
E.Verlinde and H.Verlinde \journal Phys.Lett&B192(87)95.
}
\REF\lust{
G.Lusztig \journal Adv.Math.&70(88)237.
}
\pubnum={RIMS-927}
\date={\quad }
\pubtype={\quad}
\titlepage
\title{BRST Cohomology in Quantum Affine Algebra $U_q(\widehat{sl_2})$}
\author{HITOSHI KONNO}
\address{ Research Institute for Mathematical Sciences, \break
Kyoto University, Kyoto 606-01, Japan.}
\abstract{
Using free field representation of quantum affine algebra
$U_q(\widehat{sl_2})$,
we investigate the structure of the Fock modules over $U_q(\widehat{sl_2})$.
The analisys is based on a $q$-analog of the BRST formalism
given by Bernard and Felder in the affine Kac-Moody algebra
 $\widehat {sl_2}$.
We give an  explicit construction of the singular vectors
using the  BRST charge. By the same cohomology analysis as the
classical case ($q=1$), we obtain the
irreducible highest weight representation
space as a nontrivial cohomology group.
This enables us to calculate  a trace of
the $q$-vertex operators over
this space.
}
\endpage
\sequentialequations
{\bf 1. Introduction}\par
       Recently, a
$q$-analog of the Knizhnik-Zamolodchikov ($q$-KZ) equation was discussed by
Frenkel and Reshetikhin.\refmark{\frenkel}
They argued  that  the connection matrix of the solution of $q$-KZ equation
provides
the elliptic solution of quantum Yang-Baxter
equation of the face type.
In order to construct the solutions of $q$-KZ in the
analogous way to
the Feigin-Fuchs construction in conformal field
theory, some free field representations of
$U_q(\widehat{sl_2})$
were considered by several people for an arbitrary
representation level $k$.\refmark{\shiraishi,\matsuo,\kimura} \par
       At the same time, some of the exactly solvable lattice
models in two dimension were reformulated based upon
$U_q(\widehat{sl_2})$.\refmark{\miwa}
 This formulation allows explicit evaluation of form factors
of local operators,
as a trace of certain  $q$-vertex operators
over the irreducible highest weight
representation (IHWR) space of $U_q(\widehat{ sl_2})$.
\par
       In this letter, we discuss a $q$-analog of the
bosonized version of the
Bernard-Felder's BRST
formalism given for $\widehat{sl_2}$\refmark{\bafe,\frau, \konno}
and investigate the structure of the Fock modules over
$U_q(\widehat{sl_2})$.
We give a construction of
the singular vectors and
a picture fixing procedure,
which restricts the Fock module to the one isomorphic to the
expected $q$-analog of the Wakimoto module.
We then obtain
IHWR of $U_q(\widehat{sl_2})$
as the BRST cohomology group in a  complex of  the Fock modules.
As an application, we discuss a general formula which allows
the calculation of traces of the $q$-vertex operators
over IHWR. This formula is useful for the construction
of the solutions of $q$-KZ equation (i.e. form factors)
in the
exactly solvable models.
\nextline
{\bf 2. Preliminaries}\par
      The quantum affine
algebra $U_q(\widehat{sl_2})$ is realized by the three bosonic fields
$\Phi, \phi$ and $\chi$. These fields are given by\refmark{\shiraishi}
$$\eqalignno{
X(L;M,N\vert z;\alpha) = &
- \sum_{n \neq 0}
{ [{Ln}] a_{X,n}\over
[{Mn}][{Nn}]} z^{-n}q^{|n|\alpha}
+ { L\tilde{a}_{X,0}\over MN} \log z
+ { LQ_{X}\over MN}&\eqname{\boson}
\cr}$$
for $X=\Phi,\phi$ and $ \chi$ and
$L,M,N \in \bf Z_{>0}$, $\alpha \in \bf R$.
Here as usual, $|q|<1$ and
$$\eqalignno{
[m]& = {q^{m}-q^{-m}\over q-q^{-1}},&\eqname{\qnum}\cr}
$$
for $m \in \bf Z$.
We also use the notation
$$\eqalignno{
X({N}\vert{z};{\alpha}) =&
X({L};{L},{N}|{z};{\alpha}).&\eqname{\bosona} \cr
}$$
The above three fields are quantized by imposing the
following commutation relations:
$$\eqalignno{
[a_{\Phi,n},a_{\Phi,m}]& = \delta_{n+m,0}
{ [2n][{(k+2)n}]\over n}, \quad
[\tilde{a}_{\Phi,0},Q_{\Phi}]=2(k+2),&\eqname{\bosonb} \cr
[a_{\phi,n},a_{\phi,m}]& = - \delta_{n+m,0}
{ [2n][{2n}]\over n}, \quad
[{\tilde{a}_{\phi,0},Q_{\phi}}]=-4,&\eqname{\bosonc} \cr
[{a_{\chi,n},a_{\chi,m}}]& = \delta_{n+m,0}
{ [{2n}][{2n}]\over n}, \quad
[\tilde{a}_{\chi,0},Q_{\chi}]=4,&\eqname{\bosond} \cr
}$$
with $\tilde{a}_{X,0}={ q-q^{-1}\over  2\log q} a_{X,0}$
for $X=\Phi,\phi$ and $\chi$, and others commute.
\par
Let us define
 the currents $J^{3}(z), J^{\pm}(z)$
by
$$\eqalignno{
J^{3}(z) =&_{k+2}\partial_{z} \Phi({k+2}|{q^{-2}z};{-1})
+_2\partial_z \phi({2}|{q^{-k-2}z};{-{{k}\over{2}}-1}), \cr
J^{+}(z) =&
-:\left[{_1\partial_z} \exp \left\{ -\chi({2}|{q^{-k-2}z};{0}) \right\} \right]
\exp
\left\{ -\phi({2}|{q^{-k-2}z};{1}) \right\}:,&\eqname{\currp} \cr
J^{-}(z) =&
:\left[_{k+2}\partial_z\exp \left\{
\Phi({k+2}|{q^{-2}z};{-{{k}\over{2}}-1})
+\phi({2}|{q^{-k-2}z};{- 1}) \right. \right. \cr
&\qquad  \left. \left.
+\chi({k+1};{2},{k+2}|{q^{-k-2}z};{0}) \right\}\right] \cr
&\qquad \times \exp \left\{-\Phi({k+2}|{q^{-2}z};{{k}\over{2}}+1)
+\chi({1};{2},{k+2}|{q^{-k-2}z};{0}) \right\}: . \cr
}$$
Here we use the $q$-difference operator with parameter $n \in \bf Z_{>0}$
$$\eqalignno{
{_n\partial_z}f(z)& \equiv { f(q^{n}z) - f(q^{-n}z)\over
 (q-q^{-1})z}.&\eqname{\differ}\cr}
$$
The currents in \currp\  as well as the
auxiliary fields $\psi(z)$ and $ \varphi(z)$ defined by
$$\eqalignno{
\psi(z)& = K
\exp \left\{ (q-q^{-1}) \sum^{\infty}_{k=1} J^{3}_{k}z^{-k} \right\} ,
&\eqname{\auxa} \cr
 \varphi(z)& = K^{-1}
\exp \left\{-(q-q^{-1}) \sum^{\infty}_{k=1} J^{3}_{-k}z^{k} \right\} .
&\eqname{\auxb} \cr
}$$
with
$ K= q^{\tilde{a}_{\Phi,0}+\tilde{a}_{\phi,0}}$
satisfy the relations of  $U_q(\widehat{sl_2})$
obtained by
Drinfeld\refmark{\drinfeld}:
$$\eqalignno{
[J^{3}_{n},J^{3}_{m}]& =
\delta_{n+m,0} {1 \over n}[2n]
{\gamma^{n}-\gamma^{-n}\over q-q^{-1}}, \quad [J^{3}_{n},K] = 0 , \cr
KJ^{\pm}_{n}K^{-1}& = q^{\pm2}J^{\pm}_{n} , \cr
[J^{3}_{n},J^{\pm}_{m}]& =
\pm {1\over n}[{2n}] \gamma^{\mp|n|/2}J^{\pm}_{n+m} , \cr
J^{\pm}_{n+1}J^{\pm}_{m}&-q^{\pm2}J^{\pm}_{m}J^{\pm}_{n+1} =
q^{\pm2}J^{\pm}_{n}J^{\pm}_{m+1}-J^{\pm}_{m+1}J^{\pm}_{n} ,&\eqname{\drinf} \cr
[J^{+}_{n},J^{-}_{m}]& =
{1\over q-q^{-1}}(\gamma^{(n-m)/2}\psi_{n+m} -
\gamma^{(m-n)/2}\varphi_{n+m}),\cr}$$
where  $\gamma^{\pm 1/2}\equiv q^{\pm k/2}$ is the center of the algebra.
Here we gave the mode expansions as follows.
$$\eqalignno{
 \sum_{n \in \bf Z} J^{3}_{n} z^{-n-1} =& J^{3}(z) , \qquad
 \sum_{n \in \bf Z} J^{\pm}_{n} z^{-n-1} = J^{\pm}(z) , \cr
 \sum_{n \in \bf Z} \psi_{n} z^{-n} =& \psi(z) , \qquad
 \sum_{n \in \bf Z} \varphi_{n} z^{-n} = \varphi(z).&\eqname{\modes}\cr
}$$
The standard Chevalley generators $\{ e_{i},f_{i},t_{i} \}$ are given by the
identification
$$\eqalignno{
t_{0}& = \gamma K^{-1}, \quad t_{1} = K, \quad
e_{1} = J^{+}_{0}, \quad f_{1} = J^{-}_{0}, \quad
e_{0} t_{1} = J^{-}_{1}, \quad t^{-1}_{1} f_{0} = J^{+}_{-1} .&\eqname{\chev}
\cr}$$
\nextline
{\bf 3. Fock Space}\par
      The highest weight state (HWS) $|l>$,$l=0,1,..k$
for the level $k$ representation of
$U_q(\widehat{sl_2})$ is generated by
$|l>=\lim_{z\rightarrow 0}\Phi^{(l)}_l(z)|0>.$
Here the
$q$-analog of the primary field $\Phi^{(l)}_l(z)$ is defined by
\refmark{\shiraishi,\matsuo,\kimura}
$$\eqalignno{
\Phi^{(l)}_l(z)&=\exp \lbrace \Phi(l;2,k+2|q^{k}z;{k\over2}+1)
\rbrace&\eqname{\primary}\cr}$$
and $|0>$ is the vacuum state, which is annihilated by $a_{X,n}, n\geq 0$ for
$X=\Phi,\phi$ and $\chi$.
The HWS $|l>$ satisfies the relations
$t_1|l>=q^l|l>,\quad t_0|l>=q^{k-l}|l>,\quad e_i|l>=0 $
for $i=0,1$ and $p^{L_0}|l>=p^{h_l}|l>$ with $p\equiv q^{2(k+2)}$,
 $h_l={l(l+2)\over 4( k+2)}$. Here the grading operator
 $L_0=L_0^{\Phi}+L_0^{\phi}+L_0^{\chi}$ is given by
$$\eqalignno{
L_0^{\Phi}&={\tilde a_{\Phi,0}(\tilde a_{\Phi,0}+2)\over 4(k+2)}+
\sum_{n\geq 1}{n^2\over [2n][(k+2)n]}a_{\Phi,-n}a_{\Phi,n},&\eqname{\vira}\cr
L_0^{\phi}&=-{\tilde a_{\phi,0}(\tilde a_{\phi,0}-2)\over 8}-
\sum_{n\geq 1}{n^2\over [2n]^2}a_{\phi,-n}a_{\phi,n},
&\eqname{\virb}\cr
L_0^{\chi}&={\tilde a_{\chi,0}(\tilde a_{\chi,0}+2)\over 8}+
\sum_{n\geq 1}{n^2\over [2n]^2}a_{\chi,-n}a_{\chi,n}.&\eqname{\virc}\cr
}$$
\par
      Let $F_{l,s,t}$ be  the Fock space on the vector
$|l;s,t>\equiv e^{{s\over2}Q_{\phi}+{t\over 2}Q_{\chi}}|l>$
$$\eqalignno{
F_{l,s,t}&=\Bigl\lbrace \prod a_{\Phi,-n}\prod a_{\phi,-n'}\prod a_{\chi,-n''}
|l;s,t>\Bigr\rbrace,&\eqname{\fock}\cr
}$$
where $n,n'$ and $n''$ are positive integers,
and define $F_l=\oplus_{s,t\in \bf Z}F_{l,s,t}$.
Then the highest weight $U_q(\widehat{sl_2})$ module
$$\eqalignno{
V(\lambda_l)&=U_q(\widehat {sl_2})|l>&\eqname{\slmodule}\cr
}$$
of the highest weight $\lambda_l=(k-l)\Lambda_0+l\Lambda_1$
with $\Lambda_0, \Lambda_1$ being the fundamental weight
is embedded in the Fock space $F_l$.\par
       We are interested in obtaining the
$q$-analog of the Wakimoto module, i.e. the Fock space which
coincides with the Wakimoto module for $\widehat{sl_2}$
in the limit $q\rightarrow 1$.
As is in the classical case,
our Fock space $F_l$ contains some redundancies comparing
with the expected $q$-Wakimoto module.
The origin of them is in the bosonization of
the $q$ counterpart of the first order fields $\beta$ and $\gamma$
by $\phi$ and $\chi$.\refmark{\fms} That is, roughly speaking,
in \currp, we use the fields
$\phi$ and $\chi$ only in the combination
$$\eqalignno{
\beta(z)&=-\bigl[{_1\partial}_z\xi(z)\bigr]
:\exp\lbrace-\phi(2|q^{-k-2}z;-1)
\rbrace:,&\eqname{\bbeta}\cr
\gamma(z)&=\eta(z):\exp\lbrace\phi(2|q^{-k-2}z;0)
\rbrace:,&\eqname{\bgamma}  \cr
}$$
where
$$\eqalignno{
\xi(z)&=:\exp\lbrace-\chi(2|q^{-k-2}z;0)
\rbrace:,&\eqname{\bxi}\cr
\eta(z)&=:\exp\lbrace\chi(2|q^{-k-2}z;0)
\rbrace:.&\eqname{\bbbeta}\cr
}$$
Therefore
the zero-mode of $\xi(z)$,
$\xi_0=\oint{dz\over z}\xi(z)$, is  not contained in the module
$V(\lambda_l)$.\par
       Note that, as in the classical case, we have
the following operator product expansion (OPE).
$$\eqalignno{
\xi(z)\eta(w)&=-\eta(w)\xi(z)\sim{1\over z-w}&\eqname{\xeope}\cr}
$$
and especially, $\lbrace \xi_0,\eta_0\rbrace=1$ for
 $\eta_0=\oint dz\eta(z)$.
Furthermore, we have
$$\eqalignno{
J^{3}(z)\eta(w)&=\eta(w)J^{3}(z)\sim 0,\cr
J^{-}(z)\eta(w)&=-\eta(w)J^{-}(z)\sim 0,\cr
J^{+}(z)\eta(w)&=-\eta(w)J^{+}(z)\sim {_1\partial_w}\Bigl[
{1\over z-w}:\exp \lbrace -\phi(2|q^{-k-2}w;1)\rbrace:\Bigr],
&\eqname{\opejeta}\cr
}$$
These OPEs allow $\eta_0$ to commute with all the currents.
We can therefore
restrict the Fock space $F_l$ on the kernel of $\eta_0$.
\par
      We also have a redundancy of infinite pictures, i.e.
 the redundancy in the choice of the vacuum state for $\beta$ and $\gamma$.
As in the classical case, we can fix picture by setting the constraint
$\tilde a_{\phi,0}+\tilde a_{\chi,0}=0$ on the Fock space.
This implies the constraint $s=t$ in $F_{l,s,t}$. Noting the fact that
all the currents in \currp\ commute
with $\tilde a_{\phi,0}+\tilde a_{\chi,0}$,\refmark{\shiraishi}
this constraint does not introduce any inconsistencies in our Fock space
representation.\par
      Because the above two procedure is the straightforward extension of
those used in the classical theory,\refmark{\frau,\konno,\pict} we claim that
our restricted Fock space
$\tilde F_l=\oplus_{s\in\bf Z}{\rm{Ker}}\eta_0(F_{l,s,s})$
is the $q$-Wakimoto module.\foot{
In Ref.[\matsuo], Matsuo discussed
the equivalent restriction of  Fock space
 in the different bosonization scheme.}
In the later section, we will
 evaluate the character for the irreducible highest weight representation of
$U_q(\widehat {sl_2})$ based on this restricted Fock space
and will justify this point.
\nextline
{\bf 4. Screening Charges and $q$-Vertex Operator}\par
The $q$-analog of the screening operator is given
by\refmark{\shiraishi,\matsuo}
$$\eqalignno{
S(t)&=\beta(t):\exp\lbrace
-\Phi(k+2|q^{-2}t;-{k\over 2}-1)\rbrace:&\eqname{\screen}\cr
}$$
This satisfies the following OPEs.
$$\eqalignno{
J^{3}(z)S(t)&=S(t)J^{3}(z)\sim 0,\cr
J^{+}(z)S(t)&=S(t)J^{+}(z)\sim 0,\cr
J^{-}(z)S(t)&=S(t)J^{-}(z)\sim {_{k+2}\partial_t}\Bigl[{1\over z-t}
:\exp \lbrace -\Phi(k+2|q^{-2}z;{k\over2}+1)\rbrace:\Bigr].&\eqname{\opejs}\cr
S(t)\eta(w)&=\eta(w)S(t)\cr
&\sim
{_1\partial_t}\Bigl[{1\over t-w}
:\exp \lbrace -\Phi(k+2|q^{-2}w;-{k\over2}-1)
-\phi(2|q^{-k-2}w;-1)
\rbrace:\Bigr].
\cr}$$
These OPEs allows the screening charge $\int_0^{c\infty}d_ptS(t)$
to commute with all the currents and the screening operator $S(t)$ to
commute with $\eta_0$.
Here the Jackson integral is defined by
$$
\eqalignno{
\int_0^{c\infty}d_pt f(t)&=c(1-p)\sum_{m\in \bf Z}f(cp^m)p^m,
&\eqname{\jacksona}\cr
\int_0^{c}d_pt f(t)&=c(1-p)\sum_{m\in \bf Z_{\geq 0}}f(cp^m)p^m.
&\eqname{\jacksonb}\cr}$$\par
     Now
let us discuss a $q$-analog of the
screened vertex operator.\refmark{\bafe}
The $q$-screened vertex operator $\Phi_{l,m}^{(r)}(z)
:\tilde F_{l_1}\longrightarrow \tilde F_{l_3}$
with $2r=l_1+l-l_3$
is defined by
$$\eqalignno{
\Phi_{l,m}^{(r)}(z)&=\int_{0}^{q^{-l}z}d_pt_1\int_{0}^{q^{-l}z}d_pt_2\cdots
\int_{0}^{q^{-l}z}d_pt_r\Phi^{(l)}_{m}(z)
S(t_1)\cdots S(t_r),&\eqname{\qvertex}\cr}$$
where
$$\eqalignno{
   \Phi_{m-1}^{(l)} (z)& =
   {[m-1]!\over [l]!}
   \oint{ dw_{1}\over 2\pi i}\oint {dw_{2}\over 2\pi i}\cdots\oint
{dw_{l-m+1}\over 2\pi i}\cr
   & \qquad \qquad \times
   [\cdots[~[ \Phi^{(l)}_{l}(z), J^{-}(w_{1}) ]_{q^l}, J^{-}(w_{2}) ]_{q^{l-2}}
    \cdots  J^{-}(w_{l-m+1}) ]_{q^{2m-l}}
&\eqname{\bprimary}\cr}$$
for $m=1,2,..,l$.
Here the integration regions in \qvertex\ are chosen such that the
Jackson integral of any matrix elements of
operators containing
$$\eqalignno{
\Phi^{(l)}_m(z) S(t_1)S(t_2)\cdots S(t_{j-1}){_{k+2}\partial_{t_j}}
\Bigl[{1\over w-t_j}:\exp\lbrace -\Phi(k+2|q^{-2}t_j;{k\over 2}+1)\rbrace:
\Bigr]S(t_{j+1})\cdots S(t_r),
&\eqname{\totald}\cr}$$
for $j=1,2,..,r$ and arbitrary $w$, vanishes.\par
       As the classical integral,\refmark{\dotsenko}
 one can relate the above Jackson integral
to the following ordered Jackson integral
$$\eqalignno{
{\cal J}_r(z)&=\int_0^{q^{-l}z}d_pt_1\int_0^{q^{2}t_{1}}d_pt_2
\cdots \int_0^{q^{2}t_{r-1}}d_pt_r
\Phi^{(l)}_m(z) S(t_1)S(t_2)\cdots S(t_r).&\eqname{\ordered}\cr}$$
Note that this ordered integral makes the matrix element of the
operators containing \totald\ vanish, too.
In \ordered, if the variable $t_i$ passes $t_j$,
one gets the factor ${\cal A}_s(t_i/t_j)$ defined by
$$\eqalignno{
S(t_i)S(t_j)&={\cal A}_s(t_i/t_j)S(t_j)S(t_i).&\eqname{\ssass}\cr}$$
{}From \screen, one gets
$$\eqalignno{
{\cal A}_s(t_i/t_j)&=\Bigl({t_i\over t_j}\Bigr)^{2\over k+2}
{\vartheta_1({1\over 2\pi i}\ln(q^2t_i/t_j)|\tau)\over
 \vartheta_1({1\over 2\pi i}\ln(q^{-2}t_i/t_j)|\tau)}
&\eqname{\pseudoas}\cr}$$
with $\tau\equiv {\ln p\over 2\pi i}$.
In the classical limit $q\rightarrow 1$, ${\cal A}_s $ goes to
$e^{2\pi i\over k+2}$.
Noting  the fact that the
function ${\cal A}_s(z)$
is a pseudo constant, i.e. ${\cal A}_s(p^mz)={\cal A}_s(z)$
for $m\in \bf Z$
and the definition of the Jackson integral,
one gets
$$\eqalignno{
\int_0^cd_pt {\cal A}_s(t)f(t)&={\cal A}_s(c)\int_0^cd_pt f(t).
&\eqname{\asjack}\cr}$$
Using this repeatedly, we obtain
$$\eqalignno{
\Phi^{(r)}_{l,m}(z)&=\prod_{j=1}^r{1-{\cal A}_{s}^j
\over 1-{\cal A}_s}{\cal J}_r(z),&\eqname{\relqvertex}
\cr}$$
where ${\cal A}_s\equiv {\cal A}_s(q^{2+\varepsilon})$ with
$\varepsilon>0$ being a regularization parameter.
\par
     Let us finally make a connection of our
$q$-screened vertex operator $\Phi^{(r)}_{l,m}(z)$ with  the
(type I) $q$-vertex
operator $\Phi^{\lambda_{l_3}V^{(l)}}_{\lambda_{l_1}}(z)
: \tilde F_{l_1}
\longrightarrow\tilde F_{l_3}\otimes V^{(l)}_z$ in Ref.[\miwa].
Here $V^{(l)}$ denotes the $(l+1)$-dimensional $U_q(\widehat{sl_2})$
module with basis $v^{(l)}_m,\quad m=0,1,\cdots, l$.
The $q$-vertex operator
$\Phi^{\lambda_{l_3}V^{(l)}}_{\lambda_{l_1}}(z)$ is given by the
$q-$screened vertex  as follows.
$$\eqalignno{
\Phi^{\lambda_{l_3}V^{(l)}}_{\lambda_{l_1}}(z)&=
z^{h_{l_3}-h_{l_1}}
\tilde\Phi^{\lambda_{l_3}V^{(l)}}_{\lambda_{l_1}}(z),&\eqname{\qvertexa}\cr
\tilde\Phi^{\lambda_{l_3}V^{(l)}}_{\lambda_{l_1}}(z)&=
\sum_{m=0}^l\Phi^{(r)}_{l,m}(z)\otimes v_m^{(l)}.&\eqname{\qvertexb}
\cr}$$
See Ref.[\miwa] for the detailed properties of
$\tilde\Phi^{\lambda_{l_3}V^{(l)}}_{\lambda_{l_1}}(z)$.
\nextline
{\bf 5. BRST Cohomology}
\par
     In this section, we show that our Fock modules \fock\ are reducible
for the special value of the spin $l$ due to the existence of
singular vectors. Constructing these singular vectors explicitly,
we discuss their resolution in the Fock spaces.\par
     Let us define the BRST charge $Q_n$, $n\in {\bf Z}_{>0}$ as follows.
$$\eqalignno{
Q_n&={1-{\cal A}_s^n\over 1-
{\cal A}_s}\int_{0}^{\infty}d_pt\int_{0}^{q^2t}d_pt_2\cdots
\int_{0}^{q^2t}d_pt_n
S(t)S(t_2)\cdots S(t_n).&\eqname{\brst}
\cr}$$
The BRST current
$$\eqalignno{
J_n(t)&={1-{\cal A}_s^n\over 1-{\cal A}_s}
\int_{0}^{q^2t}d_pt_2\cdots
\int_{0}^{q^2t}d_pt_n
S(t)S(t_2)\cdots S(t_n)&\eqname{\brstj}\cr}$$
is a single valued function on $\tilde F_{n,n'}\equiv\tilde F_{l_{n,n'}}$.
Here $l_{n,n'}=n-n'{P\over P'}-1$ with non-negative integers
$n,n'$ and  coprime positive integers $P,P'$
satisfying ${P\over P'}\equiv k+2$.
It has been proved by
Malikov\refmark{\malikov} that the Verma module $V(\lambda_l)$
over $U_q(\widehat{sl_2})$ are reducible if and only if
its classical counterpart is reducible, i.e. $l$ satisfies the
Kac-Kazhdan equation, $l=n-n'{P\over P'}-1$ with certain
non-negative integers $n,n'$.
In the following paragraph, we are concentrate on this case.\par
      We have
\subsection{\bf Proposition}\nextline
(i)$ Q_nQ_{P-n}=Q_{P-n}Q_n=0.$\nextline
(ii)The following infinite sequence
$$\eqalignno{
&\cdots
{\buildrel Q_{n}\over\longrightarrow}\tilde F_{-n+2P,n'}
{\buildrel Q_{P-n}\over\longrightarrow}\tilde F_{n,n'}
{\buildrel Q_{n}\over\longrightarrow}\tilde F_{-n,n'}
{\buildrel Q_{P-n}\over\longrightarrow}\tilde F_{n-2P,n'}
{\buildrel Q_{n}\over\longrightarrow}\tilde F_{-n-2P,n'}
{\buildrel Q_{P-n}\over\longrightarrow}
\cdots &\eqname{\complex}
\cr}$$
is a complex. \par
     The first statement follows from $Q_nQ_{P-n}=Q_{P-n}Q_n=Q_P$ and
${\cal A}_s^{P}=1$. Here the latter equality is proved for the cases
$k=0,1,2$ by using Riemann's theta identity.\refmark{\mum} For example,
 in the case $k=2$, the following identity is used.
$$\eqalignno{
2\vartheta_1(x|\tau)^4&=\sum_{\alpha\in all\ spin \atop structures}
e_{\alpha}\vartheta[\alpha](2x|\tau)\vartheta[\alpha](0|\tau)^3,&\eqname{\thetaid}\cr}$$
where $e_{00}=e_{10}=1=-e_{01}=-e_{11}.
$The remaining cases ($k\not=0,1,2$) are conjectured.
The second statement follows from the fact that the BRST charge
$Q_n$ commutes with $\eta_0$ and the constraint
$\tilde a_{\phi,0}+\tilde a_{\chi,0}=0$.
\par
     We also have
\subsection{\bf Lemma}\nextline
The singular vector $Q_n|n,-n'>(\not=0)$ exist for positive integers $n'$
and $n$ with $1\leq n\leq P-1$. In addition, for the same integers $n'$ and
$n$, there exists a state ( cosingular vector )
 $|w>\in \tilde F_{n,n'}$ such that $Q_n|w>=|-n,n'>$.
Here $|n,n'>\equiv |l_{n,n'}>$.\par
      The proof follows from the analogous calculation to the classical
 case.\refmark{\bafe}
$$\eqalignno{
&Q_n|n,-n'>\cr
&={1-{\cal A}_s^n\over 1-
{\cal A}_s}
\int_0^{\infty}d_pt t^{-1+n(1-n')}
 \int_0^{q^2}d_pv_2\cdots \int_0^{q^2}d_pv_n
\prod_{j=2}^nv_j^{-{l_{n,-n'}\over k+2}}
{(q^{-2}v_j;p)_{\infty}\over (q^2v_j;p)_{\infty}}\cr
&\qquad\qquad\times \prod_{2\leq i<j\leq n}v_j^{2\over k+2}
{(q^{-2}v_j/v_i;p)_{\infty}\over (q^2v_j/v_i;p)_{\infty}}
\beta(t)\prod_{j=2}^n\beta(tv_j)\cr
&\times
\exp\Bigl\lbrace \sum_{m\geq 1}{a_{\Phi,-m}\over [(k+2)m]}q^{-({k\over 2}+4)m}
t^m(1+\sum_{j=2}^nv_j^m)\Bigr\rbrace|-n,-n'>,&\eqname{\singularv}\cr
}$$
where we made a change of the variables $t_j=tv_j,\quad j=2,3,..,n$.
Note $\beta_m|l>=0$ with $\beta_m=\oint{dz\over 2\pi i}z^m\beta(z)$
for $m\geq0$ and the formula
$\int_0^{c\infty}d_ptt^l=(1-p)c^{l+1}\delta(p^{l+1}),$
where $\delta(p^{l+1})$ is a delta function vanishing
unless $l\not=-1$.\refmark{\flm}
We hence have a non-vanishing condition $n'>0$ for the
$t$ integration.
Furthermore, we can evaluate the inner product of $Q_n\vert n,-n'>$
with a covector $<\widetilde{-n,-n'}\vert(\gamma_{n'})^n$,
where $<\widetilde{-n,-n'}|-n,-n'>=1$ and $\gamma_n=\oint{dz\over 2\pi i z}
z^n\gamma(z)$.
This is possible due to
the formula established by Askey, Kadell
and Aomoto.\refmark{\askey}
$$\eqalignno{
\int_0^{q^{-l}z}&d_pt_1\int_0^{q^{2}t_{1}}d_pt_2
\cdots \int_0^{q^{2}t_{r-1}}d_pt_r
\prod_{j=1}^rt_j^{x-1}{(pt_j;p)_{\infty}\over (p^yt_j;p)_{\infty}}
\prod_{1\leq i<j\leq r}t_i^{2K-1}{(p^{1-K}t_j/t_i;p)_{\infty}\over
(p^{K}t_j/t_i;p)_{\infty}}(t_i-t_j)\cr
&=
p^A\prod_{j=1}^r{\Gamma_p(x+(r-j)K)\Gamma_p(y+(r-j)K)\Gamma_p(jK)
\over\Gamma_p(x+y+(2r-j-1)K)\Gamma_p(K)}
&\eqname{\aka}\cr}$$
for $A=\sum_{j=1}^r(x+2K(r-1)+1-j)(1+(j-1)K)$.
Here,
$(a;p)_{\infty}=\prod_{s=0}^\infty(1-ap^s)$ and
$$\eqalignno{
\Gamma_p(x)&={(p;p)_{\infty}\over(p^x;p)_{\infty}}(1-p)^{1-x}
&\eqname{\qgamma}\cr}$$
is a $q$-gamma function.\refmark{\askey}
The result is given by
$$\eqalignno{
<\widetilde {-n,-n'}|&(\gamma_{n'})^nQ_n|n,-n'>
\sim\prod_{j=1}^n{1-{\cal A}_s^j\over 1-{\cal A}_s}
\prod_{j=1}^{n-1}{\Gamma_p(-{j\over k+2})\Gamma_p({j\over k+2})\over
\Gamma_p({1\over k+2})}.
\cr}$$
This
implies the non-vanishing condition $1\leq n\leq P-1$.
The second part of the  statement follows from the same
calculation.\refmark{\bafe}
\par
     In the case
$1\leq n\leq P-1$ and $n'=1$, we obtained consistent expressions of
the singular vectors with
those obtained by Malikov\refmark{\malikov}
in the Chevalley basis.
To show this coincidence in
the other cases may be an interesting problem.\par
     This lemma as well as the proposition indicates that the same
$U_q(\widehat{sl_2})$ submodule
structure generated by the singular and cosingular
vectors as the classical case  may exist in our Fock space $\tilde F_l$.
 We thus claim that the IHWR ${\cal H}_{n,n'}$ of $U_q(\widehat{sl}_2)$
with the highest weight $\lambda_{l_{n,n'}}$ is given by the BRST
cohomology group as follows.
$$\eqalignno{
Ker Q_n^{[s]}/Im Q_n^{[s-1]}&=\Bigl\lbrace\matrix{0&for&s\not=0\cr
                                           {\cal H}_{n,n'}&for&s=0\cr},
&\eqname{\cohom}\cr}$$
where $Q^{[2a]}_n=Q_n$ and $Q_n^{[2a-1]}=Q_{P-n}$ with $a\in \bf Z$.
 The proof may be carried out in the same way as in the classical case by
applying Jantzen filtrations.\refmark{\bafe,\fefr}
\par
     As a corollary, we have a formula for the trace over the IHWR of
$U_q(\widehat{sl_2})$
$$\eqalignno{
\Tr_{{\cal H}_{n,n'}}{\cal O}&=\sum_{s\in \bf Z}(-)^s
\Tr_{{\tilde F}^{[s]}_{n,n'}}{\cal O}^{[s]},
&\eqname{\tracef}\cr}$$
where the graded physical operator ${\cal O}^{[s]}$ is defined recursively by
the relations
$$\eqalignno{
Q_n^{[s]}{\cal O}^{[s]}&={\cal O}^{[s+1]}Q_n^{[s]},\quad
{\cal O}^{[0]}={\cal O}.
&\eqname{\phys}\cr}$$
\par
      The $q$-screened vertex operator $\Phi^{(r)}_{l,m}(z)$ is a
typical physical operator. The BRST relations for this vertex are
obtained from \qvertex\ and \brst\ as follows.
$$\eqalignno{
Q_{n_3}\Phi^{(r)}_{l,m}(z)&={\cal A}_l^{n_3}
\Phi^{(r+n_3-n_1)}_{l,m}(z)Q_{n_1},&\eqname{\brstrela}\cr
Q_{P-n_3}\Phi^{(r+n_3-n_1)}_{l,m}(z)&={\cal A}_l^{P-n_3}
\Phi^{(r)}_{l,m}(z)Q_{P-n_1},&\eqname{\brstrelb}\cr
}$$
where ${\cal A}_l\equiv {\cal A}_l(q^{-l+\varepsilon})$
is defined by the relation
$S(t)\Phi^{(l)}_l(z)={\cal A}_l(t/z)\Phi^{(l)}_l(z)S(t)$.
By the explicit calculation, we have
$$\eqalignno{
{\cal A}_l(z)&=z^{-{l\over k+2}}{\vartheta_1({1\over 2\pi i}\ln(q^{-l}z)|\tau)
\over \vartheta_1({1\over 2\pi i}\ln(q^{l}z)|\tau)
}.&\eqname{\pseudoal}\cr
}$$
This is again a psuedo constant.
\nextline
{\bf 6. Trace over IHWR}\par
       We here  discuss an application of the trace formula \tracef.
 We  give an explicit picture fixing
procedure mentioned previously.\par
       The  goal is to calculate the following trace
$$\eqalignno{
\Tr_{{\cal H}_{n,n'}}&\Bigl(
\zeta^{L_0-{c\over 24}}\prod_{i=1}^N\Phi^{(r_i)}_{l_i,m_i}(z_i)
\Bigr)&\eqname{\oneloop}\cr}$$
where $c={3k\over k+2}$.
{}From the charge conservation for the fields $\Phi, \phi$ and $\chi$,
we obtain the selection rules
$$\eqalignno{
\sum_{i=1}^{N}l_i&=2\sum_{i=1}^Nr_i,\qquad
\sum_{i=1}^N m_i=\sum_{i=1}^Nr_i.&\eqname{\selection}
\cr}$$
These are the same conditions as in the classical case.
\par
       We claim that the trace over the restricted Fock space is
evaluated as
$$\eqalignno{
\Tr_{\tilde F^{[s]}_{n,n'}}{\cal O}^{[s]}&
=\Tr_{F^{[s]}_{n,n'}}\Bigl(\xi(w_0)
\oint{dw\over 2\pi i}\eta(w)
{\cal O}^{[s]}\Bigr)
\Bigl|_{\tilde a_{\phi,0}+\tilde a_{\chi,0}=0}.
&\eqname{\picturefix}\cr}$$
This is a direct extension of the classical one in Ref.[\konno].
By using this, we obtain a general formula for the trace of operators in
the $\phi\chi$ sector over the
restricted Fock space ${\tilde F}^{\phi\chi}$, where
${\tilde F}_{n,n'}={F}_{n,n'}^{\Phi}\otimes{\tilde F}^{\phi\chi}$.
The result is
$$\eqalignno{
&\Tr_{{\tilde F}^{\phi\chi}}\Bigl(\zeta^{L_0^{\phi}+L_0^{\chi}-{1\over 12}}
z^{-\tilde a_{\phi,0}}\prod_{i=1}^N:e^{-\chi(2|q^{a_i}x_i;0)}:
\prod_{j=1}^N:e^{\chi(2|q^{b_j}y_j;0)}:\prod_{r=1}^M
: e^{\phi(2|q^{c_r}z_r;d_r)}:
\prod_{s=1}^M: e^{-\phi(2|q^{g_s}w_h;h_s)}:
\Bigr)\cr
&\qquad\qquad=\eta(\tau){z^{-1}\prod_{r}Z_r\over \prod_{j=1}^NY_j
\prod_r(q^{2d_r}\zeta;\zeta)_{\infty}\prod_s(q^{2h_s}\zeta;\zeta)_{\infty}}\cr
&\qquad\qquad\times
{\prod_{0\leq i<i'\leq N} E(X_i,X_{i'})
\prod_{j<j'} E(Y_j,Y_{j'})\prod_{r,s}E_q(Z_r,W_s;d_r+h_s)
\over \prod_{0\leq i\leq N
\atop 1\leq j\leq N} E(X_i,Y_j)
\prod_{r<r'} E_q(Z_r,Z_{r'};d_r+d_{r'})\prod_{s<s'}
E_q(W_{s},W_{s'};h_s+h_{s'})}
\cr
&\qquad\qquad\times{\prod_{j'=1}^N
\vartheta_1(-Y_{j'}+\sum_{0}^NX_i-\sum_1^NY_j+\sum_1^MZ_r-\sum_1^M
W_s-z^2|\tau)
\over
\prod_{i'=0}^N\vartheta_1(-X_{i'}+
\sum_{0}^NX_i-\sum_1^NY_j+\sum_1^MZ_r-\sum_1^MW_s-z^2|\tau)}
&\eqname{\xieta}\cr}$$
where $X_i=q^{a_i}x_i, Y_j=q^{b_j}y_j, Z_r=q^{c_r}z_r, W_s=q^{g_s}w_s,
X_0=q^{-(k+2)}w_0$,
$\zeta\equiv e^{2\pi i\tau}$.\foot{Do not confuse $\tau$ in \xieta\
with the one in (36) and (53).
} The functions  $\eta(\tau)$ and $E(x,y)$
are Dedekind's function and the prime form, respectively. In \xieta, we
omitted an irrelevant factor depending only on $q$ and
used the abridged notations $x_i$ to express ${\ln x_i\over 2\pi i}$ etc.
 in the theta functions. We also introduced the
notation
$$\eqalignno{
E_q(w,z;\alpha)&={(1-q^{\alpha}z/w)\over \sqrt{z/w}}
\prod_{s=1}^{\infty}{(1-q^{\alpha}\zeta^sz/w)(1-q^{\alpha}\zeta^sw/z)}.
&\eqname{\qprimaryf}
\cr}$$
The formula \xieta\ is  a $q$-analog
 of the formula in Ref.[\verlinde] for genus one.
\par
        Using this formula,
the character ${\cal X}_{n,n'}(\tau)$ of the IHWR ${\cal H}_{n,n'}$
is evaluated as follows.
$$\eqalignno{
{\cal X}_{n,n'}(\tau)&=\Tr_{{\cal H}_{n,n'}}
\Bigl(\zeta^{L_0-{c\over 24}}z^{-J^3_0}\Bigr)\cr
&={1\over \vartheta_1(\tilde z^2|\tau)}
\sum_{s\in \bf Z}\Bigl[
 \zeta^{PP'(s-{nP'-n'P\over 2PP'})^2}{\tilde z}^{2P(s-{nP'-n'P\over 2PP'})}
- \zeta^{PP'(s+{nP'+n'P\over 2PP'})^2}
{\tilde z}^{2P(s+{nP'+n'P\over 2PP'})}\Bigr],
\cr\eqinsert{\eqname{\character}}}$$
with $\tilde z=z^{2\ln q\over q-q^{-1}}$.
This coincides with the Weyl-Kac character formula for $\widehat{sl_2}$
being  consistent with the result in the $q-$deformed Verma
module.\refmark{\lust}
\endpage
{\bf Conclusion}\par
       We discussed a $q$-analog of Bernard-Felder's BRST formalism for
$u_q(\widehat{sl_2})$.
We defined the nilpotent BRST operator $Q_n$ and
constructed the singular vectors.
Combining this with the picture fixing procedure,
we obtained the IHWR of the quantum affine algebra
$U_q(\widehat{sl_2})$ as the BRST cohomology group.
Hence a trace of the $q$-vertex operators
over IHWR was made a calculable form.
\par
       This formula allows us to
construct the  solutions of the $q$-KZ equation.
In addition, according
to Ref.[\miwa],
the evaluation of the spin-spin correlation functions
as well as the form factors in the
higher spin XXZ model is now possible.
We will discuss these subjects elsewhere.\nextline
{\bf Acknowledgement}\par
       The author would like to thank S.Helmke,
K.Kimura, A.Matsuo, A.Nakayashiki,
I.Ojima and J.Shiraishi for valuable discussions.
\refout
\end
\bye